\documentclass{bioinfo}
\copyrightyear{2017}
\pubyear{2017}

\begin{document}
\firstpage{1}

\title[Combinatorial GO score for network alignments]{SANA NetGO: A combinatorial approach to using Gene Ontology (GO) terms to score network alignments}
\author[Hayes, Mamano]{Wayne B. Hayes\footnote{to whom correspondence should be addressed ({\tt whayes@uci.edu})}\;, Nil Mamano}
\address{Department of Computer Science, University of California, Irvine CA 92697-3435, USA}

\history{Received on XXXXX; revised on XXXXX; accepted on XXXXX}

\editor{Associate Editor: XXXXXXX}

\maketitle

\begin{abstract}

Gene Ontology (GO) terms are frequently used to score alignments between protein-protein interaction (PPI) networks.  Methods already exist to measure the GO similarity between two proteins in isolation, and network alignment GO measures typically take the mean of pairwise similarities across aligned pairs of proteins.  However, pairs of proteins in a network alignment are not {\em isolated}, because each pairing is implicitly dependent upon every other pairing via the alignment itself, which is constructed globally.
Furthermore, taking a mean of pairwise scores fails to take into account the frequency of GO terms across the networks. Some GO terms are very infrequent and thus very informative to the alignment, while others are so common that even a random alignment will match a decent number of them. Existing network alignment GO scoring methods attempt to account for this in an {\it ad hoc} fashion by imposing arbitrary rules on when to ``allow'' GO terms based on their location in the GO hierarchy, rather than using readily available frequency information in the PPI networks themselves.  Here we develop a new measure, NetGO, that naturally weighs infrequent, informative GO terms more heavily than frequent, less informative GO terms, without requiring arbitrary cutoffs.  In particular, NetGO downweights the score of frequent GO terms according to their frequency in the networks being aligned. This is a {\em global} measure applicable only to alignments, independent of pairwise GO measures, in the same sense that the edge-based EC or $S^3$ scores are global measures of topological similarity independent of pairwise topological similarities. We demonstrate the superiority of NetGO by creating alignments of predetermined quality based on homologous pairs of nodes and show that NetGO correlates with alignment quality much better than any existing GO-based alignment measures.  We also demonstrate that NetGO provides a measure of taxonomic similarity between species, consistent with existing taxonomic measures---a feature not shared with existing GO-based network alignment measures.  Finally, we re-score alignments produced by almost a dozen aligners from a previous study and show that NetGO does a better job than existing measures at separating good alignments from bad ones.

\section{Contact:} \href{name@bio.com}{name@bio.com}
\end{abstract}

\section{Introduction}
In the past decade, the alignment of protein-protein interaction (PPI) networks has received much attention, with more than a dozen alignment algorithms introduced \citep{IsoRank,GRAAL,MIGRAAL,CGRAAL,HGRAAL,SPINAL,NETAL,PISWAP,GHOST,HubAlign,WAVE,GREAT,MAGNA++,CytoGEDEVO,NATALIE2,OptNetAlign,LGRAAL,MamanoHayesSANA}. The goal of such alignments is to discover ``similar'' proteins across species, in the hopes that information from better-understood proteins in one species can be transferred to less-well-studied proteins in another. Although sequence information is commonly used for this purpose \citep{ncbi2016database}, there is strong evidence that network topology also encodes significant biological information \citep{GRAAL,topologyFunctionRelationship}.

The large number of network alignment algorithms are a testament both to its perceived importance, as well as to its difficulty: network alignment is NP-complete \citep{Cook:1971:CTP:800157.805047}, being a generalization of the subgraph isomorphism problem.
Thus, approximate heuristics must be used, and the approximations produce sub-optimal solutions whose quality must be carefully assessed.  Alignments (and the algorithms used to create them) are scored in many different ways, depending upon what one wishes to emphasize.  Some scoring functions are used to {\em guide} the creation of alignments, while others are used to evaluate an alignment after-the-fact. 
Most alignment algorithms use some sort of network topology to help guide the alignment since network topology has been shown to recover such information as phylogeny \citep{GRAAL} and correlate to function \citep{topologyFunctionRelationship}.  Usually protein sequence similarity is also used to guide the alignment, in a trade-off with topological information.

The Gene Ontology \citep{GO} is a large, hierarchical corpus of descriptive terms describing various {\it biological processes (BP), cellular components (CC)}, and {\it molecular functions (MF)} that occur in a cell.
For our purposes, we note that each protein in a PPI network is typically {\it annotated} with various GO terms.
Near the top of the GO hierarchy, terms are very general (such as the BP term ``cell division'' or CC term ``nucleus'') and many (sometimes thousands) of proteins can carry those annotations. As one descends the hierarchy, terms get more specific, and fewer proteins are annotated with them.
Some proteins are more well-understood and have many GO terms, while other proteins are less-well understood and have few (or no) GO terms associated with them.  It is for this reason that network alignment is useful: a well-annotated protein in one species may be able to help annotate a less-well-understood protein in another species, {\em if} the quality of the network alignment warrants it.

Since network alignment is still in its infancy, the plethora of algorithms must be evaluated against each other based on the biological information they recover, and using GO terms after-the-fact is a common way to measure the quality of alignments.  
The hope in a network alignment is that proteins from one network are aligned to similar proteins in the other, as independently measured by GO terms.  (Of course some day we may wish to use GO terms to {\em guide} the creation of alignments, but for now their primary use is to evaluate alignment algorithms that do not use GO terms to aid creating the alignment.)

Using GO terms to evaluate the similarity between a pair of proteins is tricky for many reasons (see  \cite{pesquita2009semantic} for a comprehesive survey).  First, GO annotations of proteins are noisy, containing both false positives and false negatives. Second, quantifying the information conveyed by any one GO term in the complex hierarchy is so difficult that over a dozen methods have been proposed, with no clear winner \citep{pesquita2009semantic}. Third, the above problem becomes even more difficult if we wish to quantify the semantic similarity between {\em two} GO terms.  Fourth, the difficulty is multiplied yet again when we wish to evaluate the semantic similarity between two {\em proteins} that are each annotated with multiple GO terms. Again, many methods have been proposed, with no clear winner and no agreed-upon gold standard \citep{pesquita2009semantic}.

Finally---and this is the topic of this paper---all existing methods of evaluating protein similarity using GO terms are designed to be applied to just two proteins {\it in isolation} \citep{pesquita2009semantic}; the ``GO similarity'' of a network alignment is typically computed as the mean GO similarity across all pairs of aligned proteins. The problem with this approach is that a network alignment is not a set of independent, isolated pairs of proteins. Instead, each pair of aligned proteins is implicitly dependent on every other aligned pair via the alignment itself, which is constructed globally. No existing GO measure takes this global dependency into account. Furthermore, the high frequency (and low specificity) of some GO terms cause even random network alignments to appear high quality unless one discards these low specificity GO terms. Typically, some arbitrary cutoff is applied to discard terms that are high (close to the root) in the GO hierarchy. While such a cutoff may make sense when comparing two proteins in isolation, a network alignment provides a more natural way to discount common GO terms: we can simply scale the utility of a GO term inversely with the frequency it appears across the networks being aligned. Thus, a GO term that appears only once in each network is appropriately viewed as a strong indicator that those two proteins should be aligned as a pair; and a GO term that appears almost everywhere in both networks has its utility (for the purpose of alignment evaluation) suitably scaled down to almost zero.

The paper is organized as follows.  Section \ref{sec:GO} describes the GO system and similarity measures in more detail. Section \ref{sec:gocov} introduces our new alignment-based measure (and several minor variations).  Section \ref{sec:comparisons} presents comparisons of all the measures as a function of alignment ``quality'' based on network alignments of ``known'' quality. Finally, section \ref{sec:final} presents some discussion and conclusions.

\section{Pairwise GO scoring}
\label{sec:GO}
\begin{figure}[ht]
\centerline{\includegraphics[width=\linewidth]{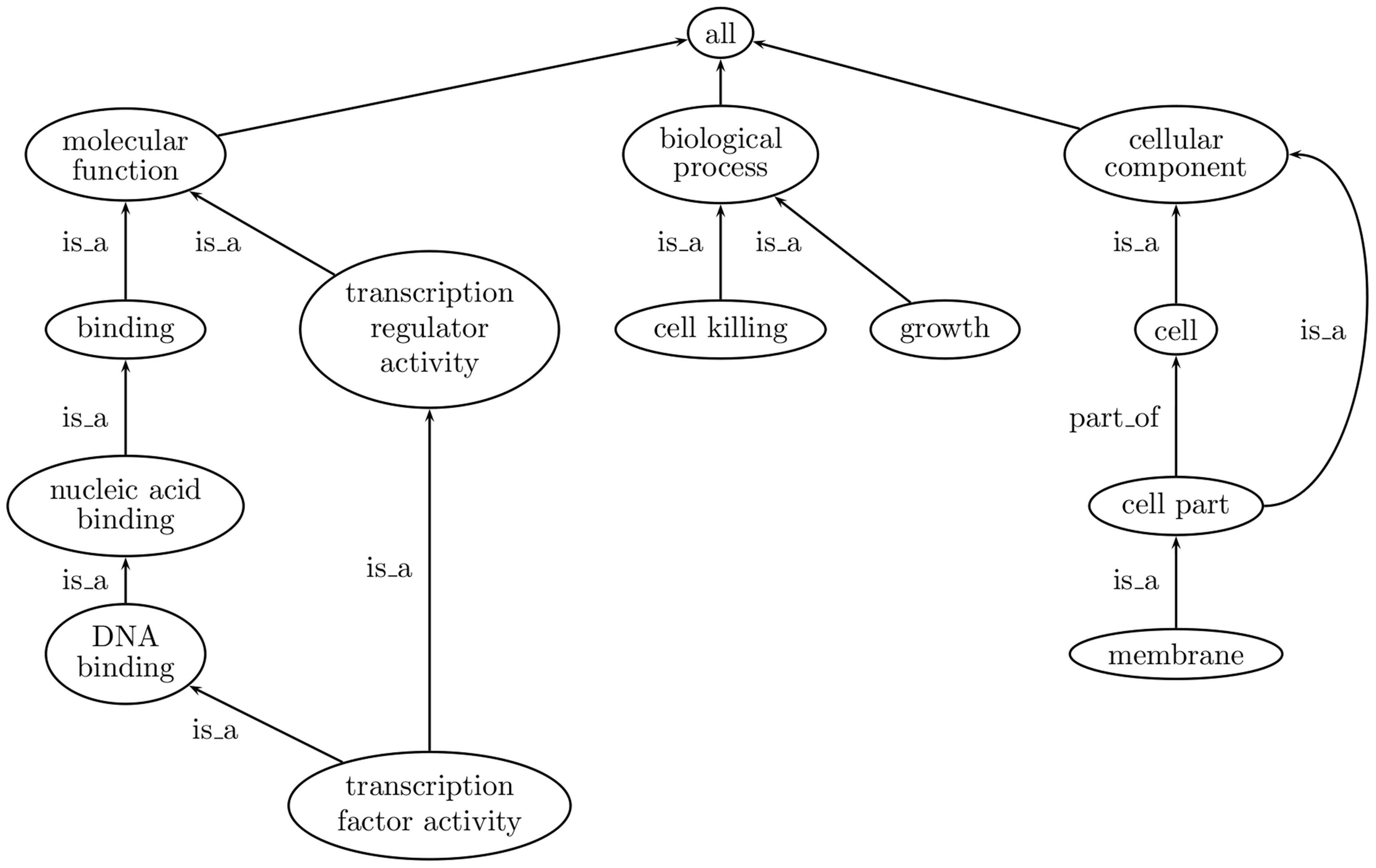}}
\caption{Sample of the GO hierarchy, taken from \cite{pesquita2009semantic}. Each node in the graph represents a GO term, with general terms nearer the top. Note that the more general a term, the greater the number of proteins are likely to be annotated with that term, and thus the {\em less} informative it is from a network alignment perspective.}
\label{fig:GO-DAG}
\end{figure}

This section draws heavily from \cite{pesquita2009semantic}, which provides an excellent and comprehensive survey of GO similarity measures.
Figure \ref{fig:GO-DAG} depicts a small portion of the GO hierarchy.
All existing measures are designed to perform pairwise comparison, either between two GO terms, or between two ``gene products'', of which proteins are an example.  To evaluate these measures we will be using the Python package {\it FastSemSim} \citep{guzzi2012semantic}, using measures listed in Table \ref{tab:semsims}.

\begin{table}[ht]
    \centering
    \begin{tabular}{|l|l|}
    \hline
         Measure & Source \\
    \hline
         Cosine & \cite{chabalier2007transversal,popescu2006fuzzy} \\
         Dice & \cite{dice1945measures,popescu2006fuzzy} \\
         Czekanowski-Dice & \cite{martin2004gotoolbox} \\
         Jaccard & \cite{popescu2006fuzzy} \\
         Jiang-Conrath & \cite{jiang1997semantic} \\
         Lin & \cite{lin1998information} \\
         NTO & \cite{mistry2008gene} \\
         Resnik & \cite{resnik1995using} \\
         SimGIC & \cite{pesquita2008metrics} \\
         SimIC & \cite{li2010effectively} \\
         SimRel & \cite{schlicker2006new} \\
         SimUI & \cite{falcon2007using}\\
    \hline
    \end{tabular}
    \caption{Measures used from the FastSemSim package.  NTO = Normalized Term Overlap.}
    \label{tab:semsims}
\end{table}

\subsection{Comparing two GO terms}
When comparing two GO terms, one can use edges and paths in the GO hierarchy to measure distance by the path length between two terms \citep{rada1989development}, or a similarity by the {\it common path} between the two terms traced back to the root \citep{wu1994verbs}.  However, both of these methods rely on the assumption that edges at the same level of the hierarchy correspond to equivalent semantic distances between terms, and that nodes at the same level have roughly equal specificity. Unfortunately, neither assumption holds in the GO hierarchy \citep{pesquita2009semantic}.

Comparing GO terms using the nodes themselves involves comparing the terms, their ancestors, or their descendants in the hierarchy.  The specificity of a particular GO term $c$ is typically measured by its {\it information content} IC($c$) = $-\log p(c)$ where $p(c)$ is the probability of $c$'s occurrence in a specific corpus (such as the UniProt database). Then, one measure of the similarity between two GO terms is the IC of their {\it most informative common ancestor} (MICA), which is their common ancestor with the highest IC (this method is the popular method by \cite{resnik1995using,resnik1999semantic}), or by the more recent {\it disjoint common ancestor} (DCA) method \citep{couto2005semantic}, which considers only common ancestors with disjoint descendant sets.

Node-based GO term comparisons are less sensitive to hierarchical assumptions than edge-based ones since they make no assumptions about level in the hierarchy.  However, they are biased by current research trends in biomedical research because terms related to current areas of scientific interest may be over-represented \citep{pesquita2009semantic}.  Still, node based measures are likely to be more useful than edge-based ones because (to directly quote \cite{pesquita2009semantic}), ``they make sense probabilistically because it is more probable (and less meaningful) that two gene products share a commonly used term than an uncommonly used term, regardless of whether that term is common because it is generic or because it is related to a hot research topic''.  In other words, IC methods appropriately down-weight common GO terms, no matter the reason they are common.

\subsection{Comparing two proteins}

Gene products such as proteins can be annotated with several GO terms, from all three of the categories MF, CC, and BP.  Thus, comparing two proteins involves comparing two sets of GO terms.  There are two major methods of comparing the two sets of GO terms: pairwise, and groupwise.  Pairwise methods look at pairs of GO terms (one from each protein), and choose either an average, or a maximum, similarity between the GO terms as representative of the similarity between the proteins.

Groupwise comparisons are broken into three categories: (i) setwise, in which set similarity (or difference) methods are used to compare the two sets of terms; (ii) graph-wise, in which subgraphs related to the two sets of terms are extracted from the GO hierarchy and then the two subgraphs are compared; and (iii) vector-wise, in which a vector space is created representing the presence or absence of terms in the two sets, and then a vector similarity function (such as dot product) is used.

\begin{methods}
\section{Scoring network alignments with GO terms}
\label{sec:gocov}

As mentioned previously, the typical method for scoring PPI network alignments using GO terms is to pick one of the many methods that score pairs of proteins, and then simply take the mean score across all pairs of aligned proteins.  Until now, this has been the only reasonable method, because all GO-scoring methods only work with pairs of individual proteins, and the mean score across the alignment seemed the only feasible way to use the existing methods.

To understand the need for a GO-based method designed specifically for evaluating network alignments, consider the following simple example.
Assume there are two networks $G_1, G_2$, each with many nodes.
Consider 4 proteins that all share a particular GO term $g$: $u_1,v_1 \in G_1$ and $u_2,v_2\in G_2$. There are two 1-to-1 alignments that could each be considered ``correct'' with respect to the GO term $g$: $(u_1,v_1)\rightarrow (u_2,v_2)$ and $(u_1,v_1)\rightarrow (v_2,u_2)$.
Since there are two equally valid alignments according to $g$, $g$ does not impose a unique alignment.
Now assume another GO term $h$ occurs only {\em once} in each network, say $w_1\in G_1$ and $w_2\in G_2$; then
there is only one ``correct'' alignment of $w_1$---it must be aligned to $w_2$. Thus, $h$ imposes a more restrictive
constraint on the aligment than does $g$, and so we consider $h$ more informative to the alignment than $g$.  Thus, aligning
$w_1$ to $w_2$ should be worth more, from the alignment perspective, than either of the two pairings of  $\{u_1,v_1\}$ to $\{u_2,v_2\}$.
In general, correctly aligning rare GO terms should be worth more than correctly aligning frequently occurring GO terms.

To make this example more concrete, consider GO terms shared between the BioGRID PPI networks of rat and mouse, including only experimentally curated terms---ie. no IEA (inferred electronically) terms.  There are exactly 87 GO terms that appear exactly once in each network, and so we can construct a unique alignment among these 87 pairs of proteins.
The mean Resnik semantic similarity score among these 87 pairs is 10.53.
Now consider GO terms that appear exactly twice in each network.
There are exactly 253 such GO terms; since each one of them annotates exactly 2 proteins in each network, each GO term introduces a 2-way ambiguity in which way we should align the 2 proteins in one network with the 2 similarly annotated proteins in the other. Across all possible alignments, the mean Resnik similarity is 10.65---slightly {\em higher} than for the uniquely defined pairings.  Although this suggests that all possible pairings carry reasonable {\em functional} similarity, our point is that in terms of {\em defining the alignment}, the doubly-occuring GO terms impose less of a constraint on the alignment and therefore those GO terms should be weighted less---not that the {\em protein} pairs should be weighted less, but the GO terms that annotate them should count for less in scoring the alignment.
(Technically, there could be as many as $2^{253}\approx 10^{76}$ ``correct'' alignments across these 253 pairs, and one can hardly claim that somehow all of these are as well-defined as the unique alignment imposed by the 87 uniquely-occurring GO terms.)
Interestingly, the two lists of proteins are not disjoint: there are 51 pairs of rat--mouse proteins that share a unique GO term and also share a (different) doubly-occurring GO term. Clearly, a unique alignment is imposed upon these 51 pairs that is not implied by the doubly-occurring GO terms, again underscoring how less-frequent GO terms are more powerful.

Of course one may argue that all of the $2^{51}$ possible above alignments convey significant value since they likely align proteins of high {\em functional} similarity even if they do not isolate homologous proteins.  This is of course correct---and in the absense of the unique proteins, the Combinatorial NetGO score (COGO in all our Figures) still would give any of the alignments about half the score of the unique ones, which is still a very respectable score reflecting high functional similarity. As such, if one is truly interested only in mean functional similarity, then certainly a ``mean semantic similarity'' score may be sufficient.

Some GO terms appear thousands of times across both networks and thus contribute virtually nothing towards constraining the alignment. In contrast, other GO terms may appear only once in each network. This extreme diversity in frequency is not sufficiently accounted for by existing schemes that down-weight common terms based only on their level in the GO hierarchy.  For example, more that 5,000 (out of about 15,000) proteins in each of rat and mouse are annotated with the GO term {\it nucleus}.  Most of these proteins are also annotated with other, less-frequent GO terms. In the context of network alignment, infrequent GO terms are far more informative than common ones. 

\subsection{NetGO: the basic idea}
In the following sections, we look at an alignment from the perspective of GO terms, rather than the perspective of proteins.  Assume a particular GO term $g$ appears $N_g$ times in one network and $M_g$ times in the other, with $N_g\leq M_g$. We say an alignment $a$ aligns the GO term $g$ ``correctly'' if all $N_g$ proteins that have it in one network are aligned with one of the $M_g$ proteins that have it in the other network.
We assign each GO term $g$ one ``unit'' of score, and spread that unit across some set of proteins or alignments, depending upon how severely we want to penalize common GO terms. We offer two methods of spreading the score below, although others could easily be concocted.

\subsection{Combinatorial NetGO score}

Given GO term $g$, we distribute the one unit of its score equally among all possible {\em alignments} that align it correctly. There are generally an exponential number of such alignments, and so this method of down-weighting common GO terms is very Draconian, exponentially down-weighting common GO terms according to their frequency. In particular, there are $P(M_g,N_g)=M_g!/(M_g-N_g)!$ ways to align $g$ correctly in the sense of having all $N_g$ proteins in one network align to the $M_g$ proteins in the other. Hence, if alignment $a$ aligns {\em all} of the $N_g$ proteins to into the corresponding $M_g$ proteins in the other network that share $g$, then $g$ contributes a total of
\begin{equation}
    1/P(M_g,N_g)
    \label{eq:P_MN}
\end{equation}
to the Combinatorial NetGO score of $a$. If $a$ only aligns $k<N_g$ of the proteins that have $g$ in one network to proteins in the other network that also have the term $g$, then $g$ contributes
\begin{equation}
    \frac{k/N_g}{P(M_g,N_g)}
    \label{eq:k_GO_P_MN}
\end{equation}
to the Combinatorial NetGO score of $a$. In other words, For GO terms that are very frequent, even aligning all of them ``correctly'' gets you almost nothing---becasue $g$ doesn't really tell you much about which {\em individual} proteins should align to each other. The total GO score of $a$ is the sum of the contribution of each GO term $g$.  Again, note that this is a sum over GO terms in the alignment, {\em not} a sum over proteins in the alignment.  This is what we mean by the scoring being GO-centered rather than protein-centered.

Finally we normalize the combinatorial GO score by dividing the sum above by the score of an alignment that, at least in principle, aligns every GO term correctly (even though such an alignment may not exist). For each GO term $g$, let $M_g$ and $N_g$ denote the number of times that $g$ appears in the two networks and assume in each case that $N_g\le M_g$.  Then, the normalization factor $Q$ is
$$Q = \sum_g\frac{1}{P(M_g,N_g)}.$$

\subsection{Inverse Frequency GO (Inverse Frequency NetGO) score}
A slightly less Draconian scoring is the {\it Inverse Frequency GO} score: given a particular GO term $g$ that appears $M_g$ times in one network and $N_g$ times in the other, we again assign exactly 1 unit of score to $g$, and assign a value of $1/\max(M_g,N_g)$ to each pair of nodes sharing $g$ that are aligned together.  
The reason we choose $\max(M_g,N_g)$ to be the denominator rather than, for example, the $\min$ or the sum, is because we want to assign the maximum score of 1 for $g$ only if the same number of nodes in both networks share $g$, and each of them is correctly aligned to a node in the other network that also has $g$. The general theme is that the only way to attain the maximum score for GO term $g$ is via a perfect alignment of all nodes in both networks that share $g$. If $N_g < M_g$ then the maximum conceivable score allotted for $g$ is $N_g/M_g$. Using this method, if the total number of distinct GO terms appearing across both networks is $K$, then we sum the scores across all $K$ GO terms $g$, and then divide by $K$, giving a total score in the range [0,1].

This second scoring method is also easily generalizable to multiple network alignment, as follows. Assume we are aligning $L$ different networks.  Given a particular GO term $g$, it will still possess exactly 1 point to be spread across the entire multi-alignment.  Let $N_{i,g}$ be the number of nodes in network $G_i$ that share $g$, and let $N'_g=\max_{i=1}^L N_{i,g}$ (the maximum number of times $g$ appears in any one network).  Then we allot a score of $1/(L\cdot N'_g)$ to each node in every network that has $g$.  However, any particular node is only {\em given} its value if there is at least one other node in its cluster that also shares $g$---and in that case all nodes in the cluster that share $g$ get to allocate their score to the total score\footnote{There are myriad other possibilities. For example perhaps the proteins should only be allowed to contribute a fraction of their value, the fraction increasing as more proteins in that cluster share $g$. Studying this issue is an area of future work.}. Just as in pairwise alignment, $g$ will obtain its maximum possible score of 1 only if every network has exactly the same number of nodes sharing $g$ (ie., $N'_g$ nodes share $g$ in every network), and every node that shares $g$ is in a cluster where {\em every} other node in the cluster also shares $g$. This would constitute a perfect alignment of all nodes that share $g$. Again, the normalizing factor (ie., denominator) will be the total number of distinct GO terms that appear across all networks.  This implies, for example, that if any network has fewer than $N'_g$ nodes that have $g$, then by the pigeon hole principle at least one cluster sharing nodes with $g$ contains at least one node that does not share $g$ with all the other nodes in said cluster, and so a perfect score with respect to $g$ is not possible.

\begin{figure*}[hbt]
\centering
\includegraphics[width=0.49\linewidth]{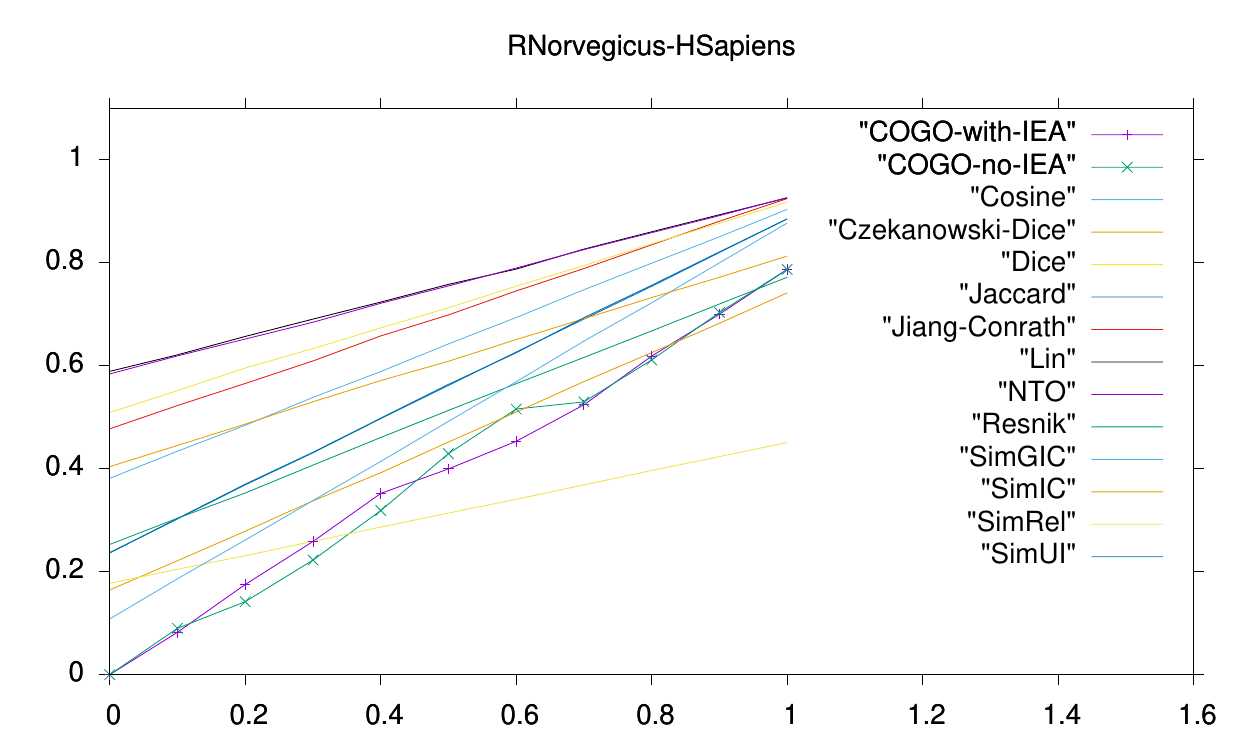}
\includegraphics[width=0.49\linewidth]{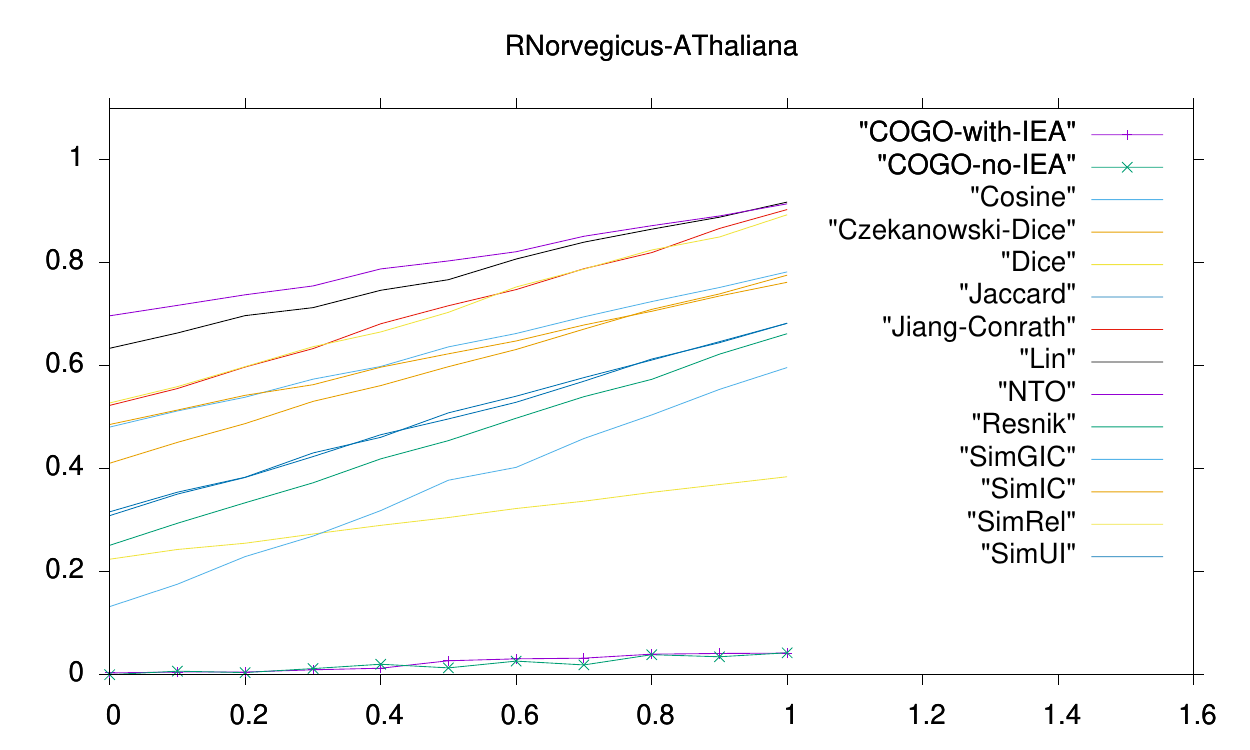}
\caption{Network alignment scores of all the methods as a function of the fraction of correctly aligned pairs of proteins, for the PPI network alignments of rat--human (left), and rat--cress (right). (See Supplement for other pairs of species.)
We note the following observations:
(i) although all of the methods produce a score that correlates with the correctness, only ours provides a score of zero when the alignment is completely randomized (ie., far left of each figure, representing alignment correctness of zero);
(ii) considering that none of the lines start at zero (other than ours), the highest scores do not sufficiently indicate how much better the best alignments are over the worst ones;
(iii) comparing the left figure to the right, and looking at the score for correctness 1 (ie., perfect alignment), none of the scores are capable of distinguishing the fact that rat--human are taxonomically much closer to each other than rat--cress, whereas our score for the perfect alignment makes this perfectly clear, giving rat--human a similarity score of 0.78, and rat--cress a score of just 0.04.  Finally, we note that our score does not change significantly depending upon whether electronically inferred (IEA) GO terms are included or not, although the latter case produces a line with slightly more noise.
(Note that Czekanowski-Dice has been reversed since it is actually a difference rather than similarity, and that Resnik has been divided by 10 since its scores tend to be in the range 3--8.)}
\label{fig:rat-scores}
\end{figure*}

\subsection{Discussion}

If there is reason to believe there is a truly ``correct'' 1-to-1 mapping of nodes, then the Combinatorial NetGO score will highly score only those alignments that have a significant fraction of the nodes aligned ``correctly''. If, on the other hand, one wishes to emphasize only functional similarity, then the Inverse Frequency NetGO score is probably more appropriate, since it will provide a good score even to common GO terms $g$, as long as all the possible nodes in one network that share $g$ are mapped to nodes in the other network that also share $g$.

One could imagine many variations on this theme.
For example, one could penalize frequent terms even less severely, such as by the logarithm of their frequency.
Or, in Equations (\ref{eq:P_MN}) or (\ref{eq:k_GO_P_MN}),
one could choose a numerator according to an existing measure of semantic similarity (such as Resnik) rather than just 1.
We have not yet explored these possibilities.

\begin{table}[hbt]
\begin{tabular}{|r|rrrrrrr|}
\hline
 &  SP & CE & MM & SC & AT & DM & HS \\
 \hline
RN & 1257  &2729 &17021 &1130  &1546  &3891  &15803 \\
SP & & 873  &1286  &1677  &806  &1056  &1284  \\
CE & &&2832  &773  &915  &2434  &2860  \\
MM &&&&1163  &1600  &4045  &16482  \\
SC &&&&&720  &948  &1164  \\
AT &&&&&&1225  &1598  \\
DM &&&&&&&4059  \\
\hline
\end{tabular}
\caption{Number of 1-to-1 (unique) homologous genes between pairs of species in the BioGRID network, according to the NCBI Homologene database. Abbreviations: RN={\it R.norvegicus}, SP={\it S.pombe}, CE={\it C.elegans}, MM={\it M.musculus}, SC={\it S.cerevisiae}, AT={\it A.thaliana}, DM={\it D.melanogaster}, HS={\it H.sapiens}.}
\label{tab:homologCounts}
\end{table}
\end{methods}

\section{Results}
\label{sec:comparisons}
To compute semantic similarities for all the measures listed in Table \ref{tab:semsims}, we use the python package FastSemSim\footnote{\tt  https://pypi.python.org/pypi/fastsemsim} \citep{guzzi2012semantic}. It implements pairwise scores between 2 proteins for the methods listed in Table \ref{tab:semsims}.

\subsection{Alignments with known correct mapping}
In order to compare our method of scoring alignments, we need to create alignments that have some sort of {\it a priori} known amount of ``correctness''.  To do this, we used the {\it NCBI homologene} database \citep{ncbi2016database}, which contains a list of known (or highly probable) homologous genes across a large array of species. Given a gene in one species, we can thus look for homologous genes in different species.  Some genes have multiple homologs in other species; for our purposes, we eliminated such pairings and allowed only 1-to-1 homologs between species so that we can uniquely create a ``correct'' alignment, even though it contains only a subset of all the proteins in each species.  Table \ref{tab:homologCounts} displays the number of 1-to-1 homologs between 8 pairs of BioGRID species, according to the Homologene database.
To create partially correct alignments, we randomly permute some fraction $F$ of those pairs; the correctness of the alignment is then defined to be $1-F$.

Figure \ref{fig:rat-scores} depicts the scores of all the methods for the two pairs of species: {\it R.norvegicus vs. H.sapiens}, and {\it R.norvegicus vs. A.thaliana}. As can be seen, all of the measures correlate with correctness.  However, none of them give an appropriately low score to alignments that are completely random. Combinatorial NetGO, on the other hand, gives such alignments a score very close to zero because the only pairs of aligned proteins that share any GO terms at all, share only the frequently occurring GO terms---the terms that are so common that even a random alignment is likely to have pairs of proteins sharing such terms.  As the correctness of the alignment increases, Combinatorial NetGO scales roughly linearly, because more homologous (and thus functionally very similar) proteins are being correctly aligned.

\begin{table*}[hbt]
\begin{tabular}{|l|lllllllllllll|}
\hline
 & COGO & SimGIC & Jaccard & SimUI & Dice & Cosine & Resnik & SimIC & SimRel & Jiang & NTO & Lin & Cz.-Dice \\
 \hline
Spearman & 0.60& 0.08& -0.01& -0.01& -0.01& -0.02& 0.08& -0.17& -0.24& -0.27& -0.37& -0.36& -0.40\\
Pearson & 0.88& 0.56& 0.46& 0.46& 0.33& 0.31& 0.25& -0.04& -0.13& -0.14& -0.21& -0.22& -0.55\\
\hline
\end{tabular}
\caption{Spearman and Pearson correlations between the taxonomic similarity, and each measure applied to the ``perfect'' alignment. Recall that the Pearson correlation measures linearity, and the Spearman measures monotonicity. Our Pearson correlation is significantly higher than all the other measures, some of which even have a negative correlation.
Our Spearman co-efficient is almsot {\em ten times} higher than the next best candidate, demonstrating that none of the measures are even close to monotonic with taxonomic similarity. Scatter plots for all measures are in the Supplementary material.}
\label{tab:pearson}
\end{table*}

\begin{figure}
    \includegraphics[width=0.49\linewidth]{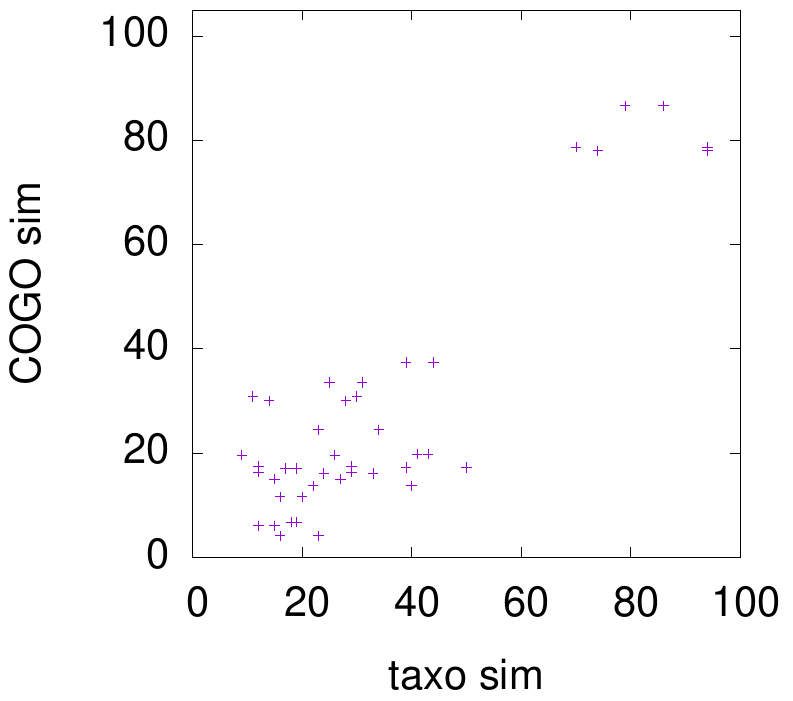}
    \includegraphics[width=0.49\linewidth]{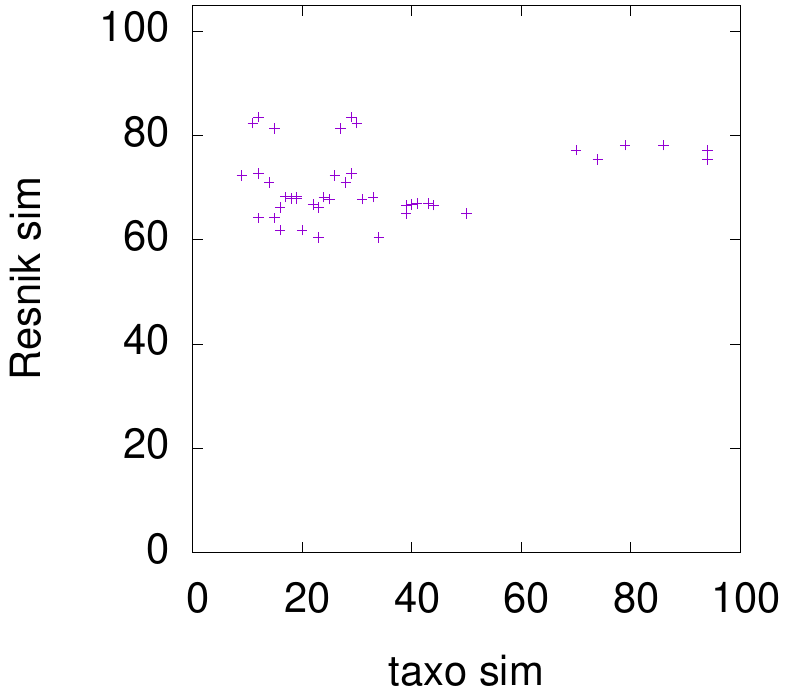}
    \caption{GO-based network scores (as a percentage) of the 28 correct alignments between pairs of networks in the BioGRID dataset, plotted against taxonomic similarity. {\bf Left}: Combinatorial NetGO score, with a Pearson correlation of 0.88 and Spearman of 0.60 with taxonomic similarity.  {\bf Right}: mean Resnik semantic similarity (averaged across all paired proteins in the alignment); Pearson correlation=0.25, Spearman=0.08. Note that taxonomic similarity is not symmetric, so each pair of species (X,Y) appears twice in this plot. The 6 points at the far right of both plots are all the combinations of the three mammals in our set (human-mouse, human-rat, mouse-rat). Note how the Resnik score does not distinguish this set from the other pairs because the network Resnik score for these 3 pairs are at the same level as the cluster of other points.}
    \label{fig:GOCO-vs-taxo}
\end{figure}

Even more interesting is that the {\em slope} of the linear correlation of Combinatorial NetGO with alignment correctness is steeper for rat--human (which are both mammals) than rat--cress (which are much more distantly related).  In fact, if we look at the Combinatorial NetGO score of the ``perfect'' alignments of both of these pairs, we find that they agree very closely to an independent measure of phylogenetic similarity \citep{gilbert2002eugenes}: both give rat--human a score in the vicinity of 0.8, and both give rat--cress a score in the vicinity of 0.05.  Figure \ref{fig:GOCO-vs-taxo} shows that this correlation extends across all 28 pairs of the 8 BioGRID species we consider.  We see that the Combinatorial NetGO score of perfect alignments has a much stronger linear correlation with taxonomic similarity than the Resnik score of those alignments, even though the Resnik score has become recently popular for scoring PPI network alignments.  In fact we see that the Resnik score is completely incapable of distinguishing mammal-mammal alignments from alignments of more distantly related species, where the Combinatorial NetGO score easily separates them.  Table \ref{tab:pearson} shows that the Combinatorial NetGO score is the only score that has a high correlation between the score and the taxonomic similarity; none of the other network alignment scores comes even remotely close, and in fact many have a zero or even negative correlation, showing that they may be inappropriate for use in a network alignment context.

\begin{figure}
\includegraphics[width=\linewidth]{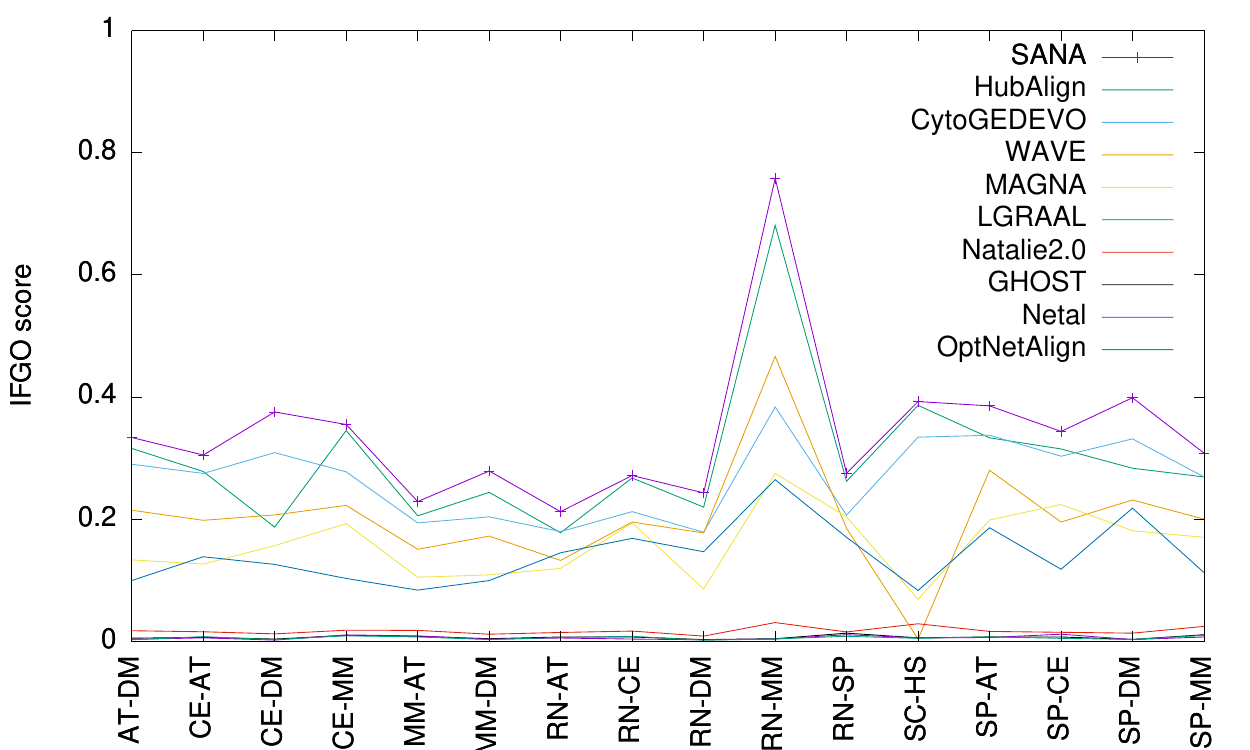}
\caption{Aligners from Mamano and Hayes (2017) evaluated with Inverse Frequency NetGO. The legend lists them in best-to-worst order according to the mean score across all 16 pairs of species.  The two-letter short names for species are the same as listed in Table \ref{tab:homologCounts}.}
\label{fig:Inverse Frequency NetGO-SANA1}
\end{figure}

\subsection{Evaluating several aligners for functional similarity}

In \cite{MamanoHayesSANA}, we evaluated alignments produced by a dozen different aligners using the Resnik semantic similarity. Among a dozen aligners, all scores were cramped in the range 2--4, and the top 5 aligners (SANA, SPINAL, CytoGEDEVO, HubAlign, and WAVE, in that order) had mean Resnik semantic similarities across 16 pairs of networks that were within 10\% of each other, with no clear-cut winner.  In Figure \ref{fig:Inverse Frequency NetGO-SANA1}, we replot the aligners using the Inverse Frequency NetGO score (which scores functional similarity rather than ``correct'' alignments).  As can be seen, SANA scores highest, although HubAlign is a close second while CytoGEDEVO, WAVE, MAGNA, and LGRAAL also have respectable scores.  All the other aligners (NATALIE, GHOST, Netal, and OptNetAlign) uniformly score almost zero.  Although the {\em ordering} of the quality of these aligners is almost identical to that presented by the Resnik score in \cite{MamanoHayesSANA}, the Inverse Frequency NetGO score much more clearly separates decent alignments from worthless ones.

\section{Discussion and Conclusions}
\label{sec:final}

We have presented NetGO, a novel method of using Gene Ontology terms to score alignments of PPI networks.  NetGO has several advantages: it is independent of the GO hierarchy, depending only upon the frequency of GO terms in the networks being aligned; it produces scores that scale well with existing methods of taxonomic similarity between species; it appropriately assigns a score near zero to random alignments, a property which no other method currently shares; it more clearly separates high from low quality alignments; and it is easily extensible to alignments between multiple PPI networks, which is a fast-growing area of biological network alignment.

NetGO is the first GO-centered measure (as opposed to protein-centered), which means that the score is computed as the sum of the contribution of each GO term, rather than the sum of the contribution of each pair of aligned proteins. This new approach to evaluating biological similarity better reflects the global nature of a network alignment, rather than considering it as a set of independent pairs of proteins.  

Similarly to how topological measures can be divided into global ones (S3, EC, WEC, etc.) and local ones (graphlet similarity, importance, etc.), so far all biological measures were strictly local. NetGO is the first global biological measure, and as in the case of topological ones, it proves to be superior to local ones in evaluating entire alignments.

%
%

\section*{Acknowledgement}
NM was supported by the Balsells fellowship.


\bibliographystyle{natbib}
\bibliography{document}
\bibliographystyle{natbib}

\end{document}